\documentclass{article}
\usepackage{epsfig}
\usepackage{cite}

\date{\today}

\usepackage{amsmath}
\usepackage{amssymb}
\usepackage[hang,nooneline,scriptsize]{subfigure}
\begin{document}

\title{\bf Inside black holes with synchronized hair}

\author{
{\large Yves Brihaye}$^{1}$,
{\large Carlos Herdeiro}$^{2}$, 
and
{\large Eugen Radu}$^{2}$
\\ 
$^{1}${\small Physique-Math\'ematique, Universite de
Mons-Hainaut, Mons, Belgium}
\\
$^{2}${\small Departamento de F\'\i sica da Universidade de Aveiro  and Centre for Research and Development in Mathematics} \\ {\small   and Applications (CIDMA), 
   Campus de Santiago, 3810-183 Aveiro, Portugal}
}

%

\date{May 2016}
\setlength{\footnotesep}{0.5\footnotesep}
\newcommand{\dd}{\mbox{d}}
\newcommand{\tr}{\mbox{tr}}
\newcommand{\la}{\lambda}
\newcommand{\ka}{\kappa}
\newcommand{\f}{\phi}
\newcommand{\vf}{\varphi}
\newcommand{\F}{\Phi}
\newcommand{\al}{\alpha}
\newcommand{\ga}{\gamma}
\newcommand{\de}{\delta}
\newcommand{\si}{\sigma}
\newcommand{\bomega}{\mbox{\boldmath $\omega$}}
\newcommand{\bsi}{\mbox{\boldmath $\sigma$}}
\newcommand{\bchi}{\mbox{\boldmath $\chi$}}
\newcommand{\bal}{\mbox{\boldmath $\alpha$}}
\newcommand{\bpsi}{\mbox{\boldmath $\psi$}}
\newcommand{\brho}{\mbox{\boldmath $\varrho$}}
\newcommand{\beps}{\mbox{\boldmath $\varepsilon$}}
\newcommand{\bxi}{\mbox{\boldmath $\xi$}}
\newcommand{\bbeta}{\mbox{\boldmath $\beta$}}
\newcommand{\ee}{\end{equation}}
\newcommand{\eea}{\end{eqnarray}}
\newcommand{\be}{\begin{equation}}
\newcommand{\bea}{\begin{eqnarray}}

\newcommand{\ii}{\mbox{i}}
\newcommand{\e}{\mbox{e}}
\newcommand{\pa}{\partial}
\newcommand{\Om}{\Omega}
\newcommand{\vep}{\varepsilon}
\newcommand{\bfph}{{\bf \phi}}
\newcommand{\lm}{\lambda}
\def\theequation{\arabic{equation}}
\renewcommand{\thefootnote}{\fnsymbol{footnote}}
\newcommand{\re}[1]{(\ref{#1})}
\newcommand{\R}{{\rm I \hspace{-0.52ex} R}}
\newcommand{\N}{{\sf N\hspace*{-1.0ex}\rule{0.15ex}%
{1.3ex}\hspace*{1.0ex}}}
\newcommand{\Q}{{\sf Q\hspace*{-1.1ex}\rule{0.15ex}%
{1.5ex}\hspace*{1.1ex}}}
\newcommand{\C}{{\sf C\hspace*{-0.9ex}\rule{0.15ex}%
{1.3ex}\hspace*{0.9ex}}}
\newcommand{\eins}{1\hspace{-0.56ex}{\rm I}}
\renewcommand{\thefootnote}{\arabic{footnote}}
 \maketitle
\begin{abstract} 
Recently, various examples of asymptotically flat, rotating black holes (BHs) with synchronized hair have been explicitly constructed, including Kerr BHs with scalar or Proca hair, and Myers-Perry BHs with scalar hair and a mass gap, showing there is a general mechanism at work. All these solutions have been found numerically, integrating the fully non-linear field equations of motion from the event horizon outwards. Here, we address the spacetime geometry of these solutions \textit{inside} the event horizon. Firstly, we provide arguments, within linear theory, that there is no regular inner horizon for these solutions. Then, we address this question fully non-linearly, using as a tractable model five dimensional, equal spinning, Myers-Perry hairy BHs. We find that, for non-extremal solutions: $(1)$ the inside spacetime geometry in the vicinity of the event horizon is smooth and the equations of motion can be integrated inwards; $(2)$ before an inner horizon is reached, the spacetime curvature grows (apparently) without bound. In all cases, our results suggest the absence of a smooth Cauchy horizon, beyond which the metric can be extended, for hairy BHs with synchronized hair.
\end{abstract}


\newpage

\section{Introduction}
The Kerr black hole (BH)~\cite{Kerr:1963ud} is the widely accepted theoretical model for describing astrophysical BH candidates~\cite{Narayan:2013gca}. It is, moreover, consistent with the recently detected transient gravitational wave event, GW150914~\cite{Abbott:2016blz}, interpreted as originating from the plunge of two BHs into a final Kerr BH, the ringdown of the latter~\cite{Berti:2009kk} being consistent with the observed signal. Still, the exact Kerr solution can only be taken as an accurate description of the physical world, outside the event horizon. Inside this horizon, the (``eternal") solution presents exotic features, including a ring singularity that allows test particles to propagate to another asymptotically flat region, wherein closed timelike curves exist~\cite{Carter:1968rr,Townsend:1997ku}. These unphysical features occur inside an internal BH horizon, which is also a Cauchy horizon~\cite{Hawking:1967ju}. This internal horizon  bounds the region of validity of the Cauchy problem, defined by initial data on any appropriate partial Cauchy surface, leading to yet more issues, this time related to the Strong Cosmic Censorship Hypothesis.

As it turns out, traveling towards this Cauchy horizon is a remarkably dangerous undertaking. Taking the spinless Kerr-Newman BH as a toy-model, Simpson and Penrose~\cite{Simpson:1973ua} observed that test, freely falling objects experience an infinite blueshift when approaching the Cauchy horizon, with respect to an asymptotic observer. They argued, moreover, that a test electromagnetic field diverges at the Cauchy horizon (see also~\cite{Chandrasekhar:1982}). This evidence suggests  an instability of the Cauchy horizon, when perturbed. The endpoint of the non-linear evolution of such instability was conjectured to be the formation of a curvature singularity, in the neighborhood of the would-be Cauchy horizon. A concrete picture for this non-linear process was put forward by Israel and Poisson~\cite{Poisson:1989zz,Poisson:1990eh}, who argued that the evolution of this instability produces a ``mass inflation", a phenomenon during which gauge invariant quantities such as  the Misner-Sharp mass~\cite{Misner:1964je}, as well as curvature invariants, grow exponentially -- see, $e.g.$,~\cite{Ori:1991zz,Breitenlohner:1997hm,Hod:1998gy,Hamilton:2008zz,Avelino:2009vv,Avelino:2011ee,Hwang:2011kg,Dafermos:2012np,Costa:2014aia,Avelino:2015fve} for discussions of this phenomenon and its endpoint. These investigations suggest that, within \textit{classical} General Relativity -- thus neglecting quantum considerations --, the internal structure of the Kerr  BH perturbed by \textit{any kind of matter/energy}, and thus the internal structure of astrophysically realistic BHs formed as a result of gravitational collapse, will be remarkably different from that of the exact (``eternal") BH solution. In particular, it will not be extendable in any physical sense beyond a would-be Cauchy horizon. 

\bigskip

Over the last two years, new families of BH solutions  that bifurcate from the vacuum Kerr BH, have been found in Einstein's gravity minimally coupled to simple matter fields, such as free, massive scalar or vector fields. These matter fields obey all energy conditions and the new families are Kerr BHs with scalar hair~\cite{Herdeiro:2014goa,Herdeiro:2015gia}, that may include self-interactions~\cite{Kleihaus:2015iea,Herdeiro:2015tia}, and Kerr BHs with Proca hair~\cite{Herdeiro:2016tmi}. These BHs can exhibit distinct phenomenological properties, $e.g.$, their shadows~\cite{Cunha:2015yba,Cunha:2016bpi,Johannsen:2016uoh}, and provide new examples of deformations of the Kerr solution, within sound and simple General Relativity models.

A common underlying feature to all these new families of solutions, is that the matter field is preserved along the orbits of the horizon null geodesic generators, but it is not preserved, independently, by the stationarity and axi-symmetry Killing vector fields.  This property can be understood as a synchronized rotation of the matter field (in terms of its phase velocity)  with the BH horizon, at the horizon~\cite{Benone:2014ssa}. Hence, we dub these BHs as \textit{synchronized hairy BHs}. Asymptotically flat sychronized Myers-Perry (MP) BHs with scalar hair (and a mass gap) were also constructed~\cite{Brihaye:2014nba,Herdeiro:2015kha} (see also~\cite{Dias:2011at} for an asymptotically Anti-de-Sitter example). All these solutions have been obtained numerically, by integrating the fully non-linear Einstein-matter field equations of motion from the horizon to infinity, and using appropriate boundary conditions.

\bigskip

Since,  at least in a large region of the parameter space, 
 synchronized hairy Kerr BHs can be seen as (non-linearly) perturbed vacuum Kerr BHs, due to the matter field, the aforementioned arguments suggest the absence of a smooth Cauchy horizon, inside the event horizon. The purpose of this paper is to initiate the investigation of this issue, which, simultaneously, is a necessary step to unveil the global structure of these solutions. 

Given the complexity of the equations of motion in the four dimensional models, here we will analyse the simpler five dimensional MP case with two equal angular momenta~\cite{Brihaye:2014nba}, for which the problem becomes co-dimension one and the equations of motion reduce to a set of non-linear, coupled ordinary (rather than partial) differential equations. Analysing this case, we provide evidence that, in accordance with the expectation built above, there is no regular Cauchy horizon for hairy BHs with synchronized hair.  Previous results on the internal structure of other types of hairy BHs can be found in, 
$e.g.$, ~\cite{Donets:1996ja,Breitenlohner:1997hm,Sarbach:1997us,Tamaki:2001wca}.

\bigskip

This paper is organized as follows. In Section~\ref{sec2} we consider a linear analysis; namely, we integrate inside the horizon the scalar field equation for the stationary scalar clouds discussed in~\cite{Hod:2012px,Hod:2013zza,Hod:2014baa,Hod:2014sha,Hod:2015ota,Hod:2015goa} (see also~\cite{Herdeiro:2014goa,Benone:2014ssa}). These are test field configuration on a vacuum Kerr BH spacetime, that can be regarded as linearized hair. We observe strong spatial oscillations near the Cauchy horizon, hence strong gradients, suggesting a high-energy feedback upon considering their backreaction. This behaviour is avoided near the outer horizon due to the synchronization condition, but such synchronization cannot be imposed at \textit{both} the inner and outer horizon, for non-extremal BHs. We also consider analogous stationary clouds around MP BHs (in a cavity), which builds a parallelism with the Kerr case. In Section~\ref{sec3} we prepare the ground for the fully non-linear analysis, by reviewing the vacuum MP BH~\cite{Myers:1986un} and the exterior hairy MP solution~\cite{Brihaye:2014nba}, for the case  of two equal angular momenta. Then, in Section~\ref{sec4} we integrate inwards the complete set of field equations for two illustrative cases of the latter solutions, and display the behaviour of the metric functions, scalar field and curvature scalar. We exhibit evidence for the inexistence of a smooth Cauchy horizon in both cases, even though the detailed behaviour is different in the two illustrative examples. Final remarks are made in Section~\ref{sec5}.

\section{Linear Analysis}
\label{sec2}

\subsection{Stationary scalar clouds on Kerr}
Kerr BHs with scalar hair~\cite{Herdeiro:2014goa,Herdeiro:2015gia} are solutions to Einstein's gravity minimally coupled to a free, massive (mass $\mu$), complex scalar field. This model is described by the action:
 \begin{equation}
\label{actionscalar}
\mathcal{S}=\int  d^4x \sqrt{-g}\left[ \frac{R}{16\pi G}
   -\frac{g^{\alpha\beta}}{2} \left( \Psi_{, \, \alpha}^* \Psi_{, \, \beta} + \Psi _
{, \, \beta}^* \Psi _{, \, \alpha} \right) - \mu^2 \Psi^*\Psi
 \right]  \ .
\end{equation}
In the vacuum Kerr limit, one can linearise the scalar field about the vacuum Kerr solution, which amounts to considering the Klein-Gordon field as a test field on this background. One can then obtain bound states of the Klein-Gordon field, dubbed stationary scalar clouds~\cite{Hod:2012px,Hod:2013zza,Hod:2014baa,Hod:2014sha,Hod:2015ota,Hod:2015goa,Hod:2016dkn,Herdeiro:2014goa,Benone:2014ssa}. Here, we shall analyse the behaviour of these (non-backreacting) clouds inside the event horizon, and in particular at the Cauchy horizon.

Taking a Kerr solution with mass $M$ and angular momentum $J\equiv a M$,  in Boyer-Lindquist coordinates $(t,r,\theta,\varphi)$, the Klein-Gordon equation admits separation of variables with the ansatz~\cite{Brill:1972xj}
%
 $\Psi=  e^{-i w t} e^{im\varphi} S_{\ell m} (\theta)R_{\ell m} (r)$,
leading to the two ordinary differential equations:
\begin{equation}
\frac{1}{ \sin\theta}\frac{d}{d \theta}\left(\sin\theta \frac{d S_{\ell m}(\theta)}{d \theta}\right)+\left[a^2\cos^2\theta(w^2-\mu^2)-\frac{m^2}{\sin^2\theta}+\Lambda_{\ell m}\right]S_{\ell m}(\theta)=0 \ ,
\label{spheroidalh}
\end{equation}
\begin{equation}
\frac{d}{d r}\left(\Delta\frac{d R_{\ell m}(r)}{d r}\right)+\left[\frac{[w(r^2+a^2)-am]^2}{\Delta}-w^2a^2-\mu^2r^2+2maw-\Lambda_{\ell m} \right]R_{\ell m}(r) =0 \ . 
\label{radialeq}
\end{equation}
The first equation defines the \textit{spheroidal harmonics} $S_{\ell m}$ (see $e.g.$~\cite{Seidel:1988ue}), where $-\ell\leqslant m\leqslant \ell$; these reduce to the familiar associated Legendre polynomials when $a=0$. $\Lambda_{\ell m}$ is a separation constant, which reduces to the familiar total angular momentum eigenvalue $\ell(\ell+1)$ in the same limit. To address the radial function $R_{\ell m}$, we recall that $\Delta$ vanishes at the horizons
\begin{eqnarray} 
\Delta=(r-r_+)(r-r_-)\ ,
\end{eqnarray}
where
\begin{eqnarray} 
r_{\pm}=M\pm \sqrt{M^2-a^2} \ ,
\end{eqnarray}
denote the outer (event) and inner (Cauchy) horizon radial coordinates, respectively.  Stationary scalar clouds are solutions of the radial equation \eqref{radialeq} possessing a critical (synchronization) frequency 
\begin{equation}
\label{cond}
\omega \equiv m\Omega_H =m\frac{a}{r_+^2+a^2} \ .
\end{equation}
They are regular at the outer horizon, $r=r_+$, and decay exponentially at spatial infinity~\cite{Benone:2014ssa}. Stationary scalar clouds form a discrete set labelled by three `quantum' numbers, $(n,\ell,m)$, which are subjected to a quantization condition, involving the BH mass. The label $n$ is a non-negative integer, corresponding to the node number of $R_{\ell m}$.  Fixing $(n,\ell,m)$, the quantization condition will yield one (physical) possible value of the BH mass. 

To address the behaviour of $R_{\ell m}$ near the horizons  we start by defining a new radial coordinate, $x$:
\begin{equation}
\Delta \frac{d}{dr}\equiv \frac{d}{dx} \ .
\end{equation}
Focusing, for now, on the exterior BH region, $i.e.$, for $ r>r_+$, we have (up to a constant)
\begin{equation}
x=\frac{1}{r_+-r_-}  \log \left(\frac{r-r_+}{r-r_-} \right) \  ,
\end{equation}
and $x\rightarrow -\infty$ as $r\rightarrow r_+$.
In this limit, the radial eq.~(\ref{radialeq})
becomes, to leading order
\begin{equation}
\label{eqx}
\frac{d^2}{dx^2}R_{\ell m}+K^2_0 R_{\ell m}=0\ ,
\end{equation}
with 
\begin{equation}
 K_0= (r_+^2+a^2)w-am \ .
\end{equation}
Thus, the solutions of eq. (\ref{eqx}) valid as $r\to r_+$ are
\begin{equation}
R_{\ell m} \sim   e^{\pm i K_0 x }  \ ,
\end{equation}
and, as long as the effective wave-vector is non-zero, $K_0\neq 0$, the radial function oscillates strongly towards the horizon. Imposing the synchronization condition~\eqref{cond}, precisely amounts to take $K_0=0$ and that allows the existence of stationary \textit{non-oscillating} solutions at the event horizon, without strong spatial gradients. Such solution can be written, close to the horizon, as a power series (eq. (\ref{eqx}) is not enough to obtain this expansion, since higher order terms in $x$ are also relevant)
	\begin{equation}
	\label{sol1}
 R_{\ell m}=r_0+r_1(r-r_+) +r_2(r-r_+)^2+\dots~ \ ,
\end{equation}
where $r_i$ are some coefficients  which are too involved to include here. One could, alternatively, postulate from the very beginning the existence of a power series solution of the form 
(\ref{sol1}). This implies directly the resonance condition (\ref{cond}).

A similar analysis can be made near the inner (Cauchy) horizon. Having imposed the synchronization condition for the outer horizon, however, there is no longer freedom to impose it \textit{also} at the inner horizon, since the horizon velocity there is different from $\Omega_H$. 
One is then left with large spatial gradients associated to the strong oscillations discussed before, which are then unavoidable at the Cauchy horizon. In Fig.~\ref{Kerr} we display the radial profile for the $\ell=1=m$ radial function (with no nodes). The oscillations are clearly seen. This is the generic behaviour, even though the integration very close to $r_-$ 
becomes increasingly difficult in terms of the $r-$coordinate (see, however, footnote 1 below). 
We have also concluded from our numerical experiments, that  observing these oscillations is easier (more difficult) for low (high) temperature BHs, 
$i.e.$ closer to (further away from) extremality.

\begin{figure}[h!]
\begin{center}
{\includegraphics[width=8.4cm]{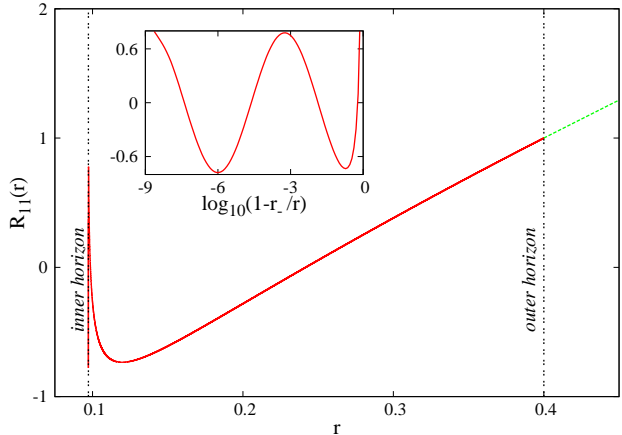}} 
\end{center}
\caption{Behaviour of  the $\ell=m=1$ nodeless radial function for a scalar field on a Kerr background with $r_H=0.4$
and $a=  0.1973$.}
\label{Kerr}
\end{figure}

\subsection{Stationary scalar clouds on $D=5$, equal angular momenta MP (in a cavity)}
The behaviour just described for the stationary scalar clouds on Kerr, $i.e$ 
smoothness on the event horizon and strong oscillations near the Cauchy horizon, should be generic for any test field obeying the synchronization condition (on the event horizon) and integrated inwards, on a BH background that possesses also an inner (Cauchy) horizon. In this subsection we support this claim by considering a massless scalar field on the $D=5$, asymptotically flat, MP BH with equal angular momentum.  We remark that, on this background, there are no exponentially decaying (towards spatial infinity) stationary scalar clouds, as a consequence of the faster fall off of the gravitational interaction in $D=5$. Indeed, the existence of hairy MP BHs is fundamentally non-linear and hence there is a mass gap with respect to the vacuum MP solutions~\cite{Brihaye:2014nba}. In the linear analysis we perform below, however, we are not interested in the asymptotic behaviour of the scalar field, towards spatial infinity. Rather, we simply want to make the point that, just as for the Kerr case in the previous subsection, a stationary, non-trivial, test scalar field that is made smooth on the outer horizon, due to the synchronization condition, will exhibit oscillations and sharp spatial gradients near the Cauchy horizon.

We consider the Klein-Gordon equation
\begin{eqnarray}
(\nabla^2 -\mu^2)\Pi=0 \ ,
\end{eqnarray}
for a complex scalar doublet, 
 $\Pi$,
on the $D=5$, asymptotically flat, MP BH with equal angular momenta. The explicit form of the scalar field and the geometry can be found in equations (\ref{scalar_ansatz}) and (\ref{metric}), (\ref{MP}), below. 
For a MP geometry, given the (outer) event horizon radius, $r_H=r_+$,
and horizon angular velocity $\Omega_H$,
an inner (Cauchy) horizon exists for
\begin{eqnarray}
r_-=\frac{r_H^2 \Omega_H}{\sqrt{1-r_H^2 \Omega_H^2}}>0\ ,
\end{eqnarray}
the extremal limit being approached as $r_H \Omega_H\to 1/\sqrt{2}$.

With this framework, the Klein-Gordon equation  reduces to
\begin{eqnarray}
\label{radialeq1}
\frac{1}{r^3}\frac{d}{dr}\left[r^3f(r)\frac{d\phi(r)}{dr}\right]+\left\{
\frac{h(r)}{ f(r)}[\omega-W(r)]^2-\left[\mu^2+\frac{1}{r^2}(2+\frac{1}{h(r)})\right] 
                                     \right\}\phi(r)=0~,
\end{eqnarray}
 where $\phi(r)$ is the scalar doublet amplitude, $cf.$ eq.~\eqref{scalar_ansatz},
$\omega$ its frequency and 
$f(r),h(r)$ and $W(r)$ are the metric functions which enter the MP line element, as given by (\ref{MP}).

A similar analysis to the one in the previous subsection can now be performed, 
\textit{mutatis mutandis}.  First, one defines a new radial variable
via the transformation
\begin{equation}
f \frac{d}{dr}\equiv \frac{d}{dx} \ .
\end{equation}
Then, as $r\to r_{c}$ (with $r_c=r_{(+,-)}$),
the radial eq.~(\ref{radialeq1})
becomes, to leading order
\begin{equation}
\label{eqx1}
\frac{d^2\phi(x)}{dx^2}+K^2_0 \phi(x)=0\ ,
\end{equation}
where, this time,
\begin{equation}
\label{MP-osc2}
K^2_0=\lim_{r\to r_c}h(r)\left[\omega-W(r)\right]^2 , 
\end{equation}
which is strictly positive, except in the synchronized case
$\omega=\Omega_H$.
Thus, the generic behaviour of the field as $r\to r_{c}$
is oscillatory.

As explained above, there are no asymptotically exponentially decaying scalar clouds (even considering a mass term for the scalar field) in this problem. But that does not concern us; 
we only
wish to investigate the behaviour inside the horizon. 
Still, to have a complete picture of the scalar field, including that outside the horizon, we shall consider our scalar field \textit{in a cavity}, $i.e.$ imposing a mirror boundary condition at some radius
\begin{equation}
r=r_0>r_H\ .
\end{equation}
The scalar
field $\phi$ is required to vanish at the cavity's boundary,
$\phi(r_0)=0$\ .
At the outer event horizon, $r=r_H$, we impose the synchronization
condition
\begin{equation}
\omega=\Omega_H\ ,
\end{equation}
such that the scalar field is smooth there,
with an approximate solution
\begin{equation}
\phi(r)=b+r_1(r-r_H)+\mathcal{O}(r-r_H)^2\ ,
\end{equation}
with 
\begin{equation}
 r_1=\frac{(1-r_H^2\omega^2)[3+r_H^2(\mu^2-\omega^2)]}{2r_H(1-2r_H^2 \omega^2) }\ .
\end{equation}
With this initial data, one integrates inwards 
the radial equation (\ref{radialeq1})
towards $r=r_{-}$.

The results of the numerical integration are similar to 
the Kerr case, even if it is harder with this radial coordinate 
to observe the oscillations,
since they occur at the very short scale close to the inner horizon.\footnote{
We have found that the oscillations can be captured by
performing the numerical integration in terms of 
a  new coordinate $\bar r$,
with $r=r_-+(r_+ - r_-)e^{-\bar r}$,
such that the outer event horizon is located at $ \bar r=0$
and the inner one is approached as $ \bar r\to \infty$.
However, this approach cannot be implemented in a non-linear setup,
since, therein, $r_-$ is not known \textit{a priori}.
} Again, far away from extremality,
it is rather difficult to find evidence for the oscillatory
behaviour predicted by the relation (\ref{eqx1}).
For lower temperatures, however,  one starts to see the predicted
oscillations -- Fig. \ref{MPfig} (right panel).

\begin{figure}[h]
\begin{center}
{\includegraphics[width=8.4cm]{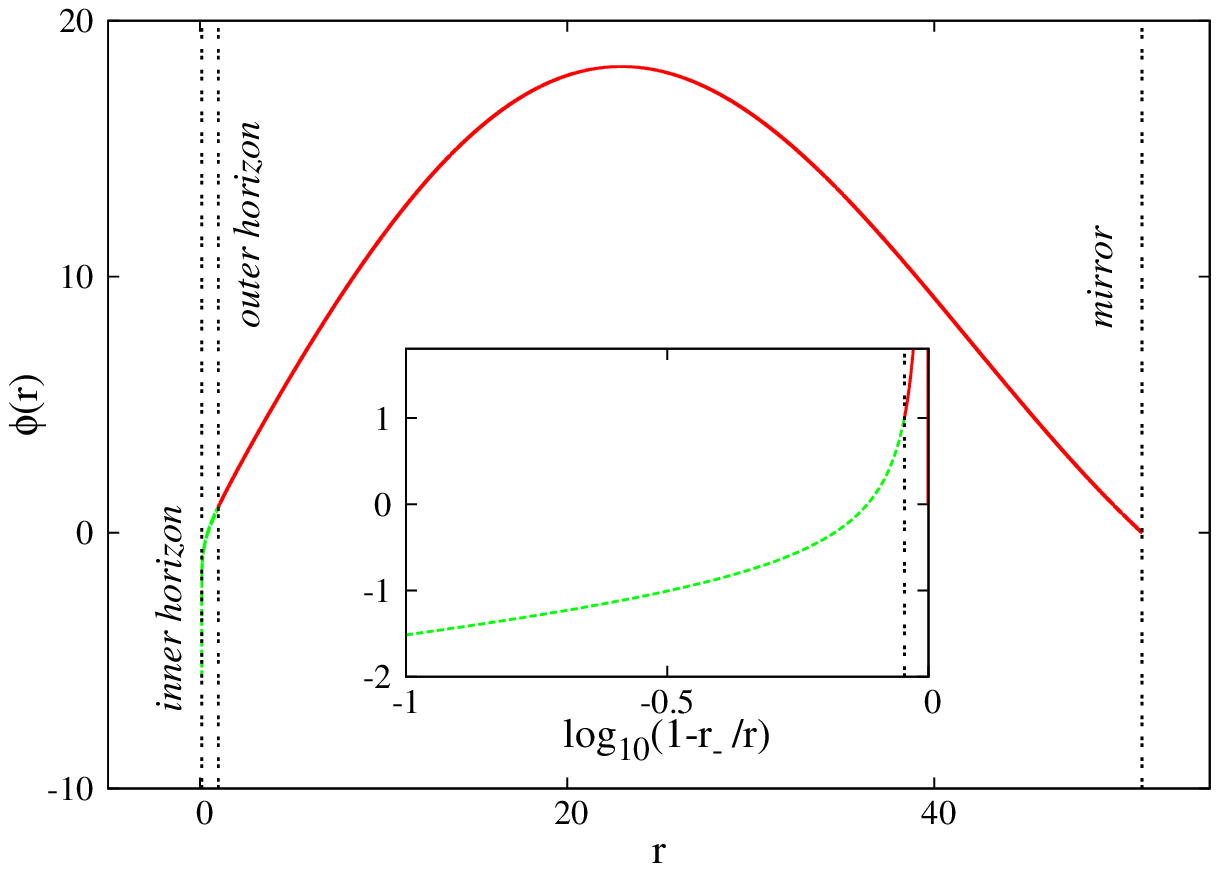}}
{\includegraphics[width=8.4cm]{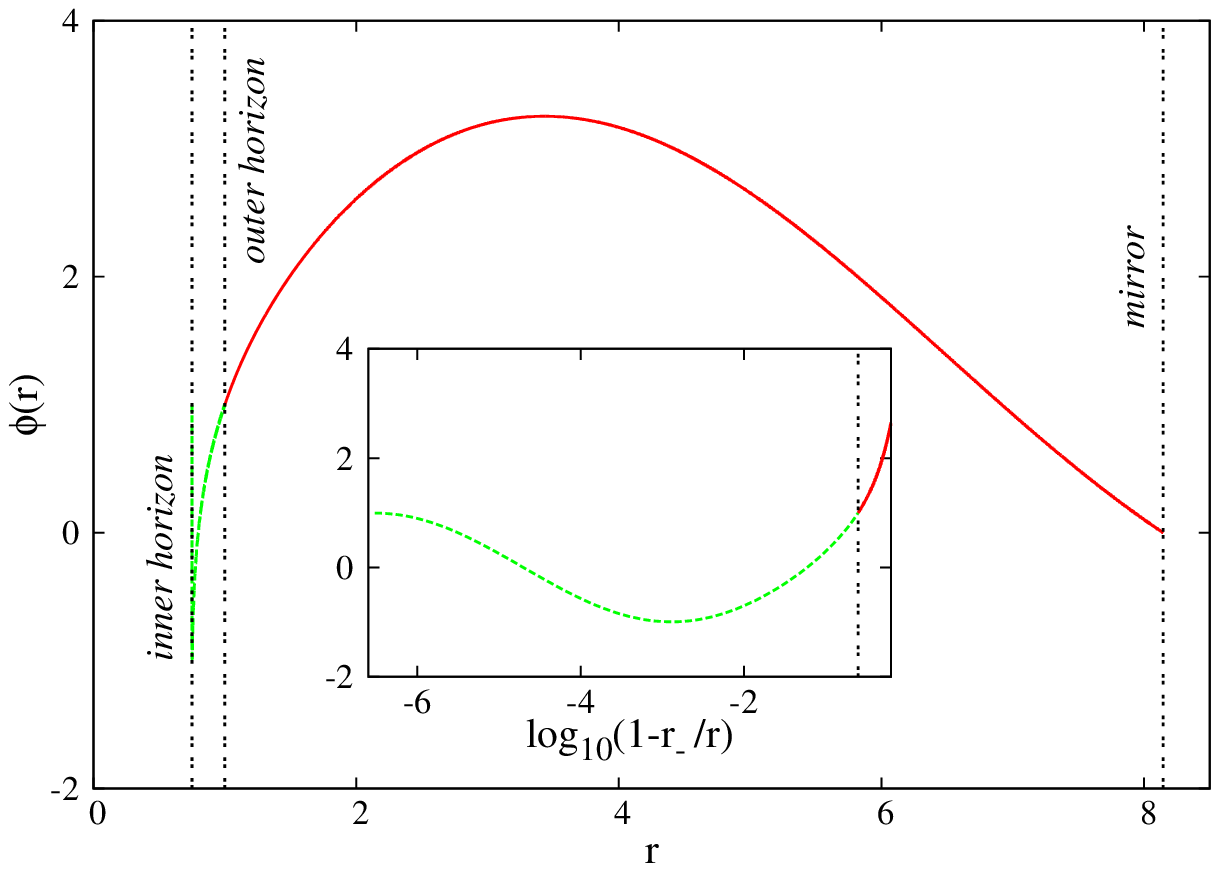}} 
\end{center}
\caption{Behaviour of the massless scalar field on two different MP backgrounds: a BH, with $r_H=1$
and $\Omega_H=0.1$ (left panel) and $\Omega_H=0.6$ (right panel). Oscillations become visible in the latter, which is closer to extremality.}
\label{MPfig}
\end{figure}  

The analysis of these subsections served to argued, from the behaviour of test scalar fields, that: 
i) the backreaction of the scalar field is likely to cause strong gravity effects, due to the strong oscillations, near the Cauchy horizon; 
ii) the behaviour is analogous for both the Kerr and the MP background. In the next section we will investigate fully non-linearly what occurs in the latter case.

\section{Non-linear Analysis: setup}
\label{sec3}
We consider Einstein's  gravity in five spacetime dimensions, minimally coupled to a massive complex scalar field doublet. 
The model is described by the following Lagrangian density
\be
\label{egbbs}
   \mathcal{S} =  \int d^5 x \sqrt{- g} \left\{\frac{R }{16 \pi G} 
   -  \partial_\nu \Pi^{\dagger} \partial^\nu \Pi -\mu^2 \Pi^{\dagger} \Pi  \right\} \ ,            
\ee
where $R$ represents the Ricci scalar and $\mu$ is the (equal) mass of the doublet of complex scalar fields, $\Pi=(\Psi_1,\Psi_2)$. The model admits a $U(2)$ global symmetry.  The Noether current associated to the $U(1)$ subgroup is 
\be
j^{\mu} = -i[\Pi^{\dagger} (\partial^{\mu}\Pi) -  (\partial^{\mu}\Pi^{\dagger}) \Pi)] \ ,
\ee 
and the corresponding
conserved charge is denoted $Q$. The variation of the action (\ref{egbbs}) with respect to the metric and the scalar field leads to the Einstein-Klein-Gordon equations, which can be found, for instance, in \cite{Brihaye:2014nba} with the same notation. 

Amongst the known solutions to the model~\eqref{egbbs} we shall be interested in stationary, rotating spacetimes admitting a $U(2)$ spatial isometry group, which include the three qualitatively distinct following cases:
\begin{description}
\item[i)] equal angular momenta, $D=5$ MP BHs~\cite{Myers:1986un};
\item[ii)] equal angular momenta, rotating boson stars. (We recall the boson stars are everywhere regular, gravitating soliton-like solutions~\cite{Mielke:1997re};
originally found in four spacetime dimensions, they have been generalized to higher dimensions by  various authors, namely in~\cite{Astefanesei:2003qy,Prikas:2004fx,Hartmann:2013tca}.) 
\item[iii)] equal angular momenta MP BHs with scalar hair (and a mass gap), found in~\cite{Brihaye:2014nba}. 
\end{description}
Whereas solutions $i)$ are known analytically in closed form, solutions $ii)$ and $iii)$ have been obtained numerically. In particular, so far, solutions $iii)$ were only constructed on and outside the event horizon. 
In the following, our goal will be to
 investigate the spacetime structure of solutions $iii)$ inside the horizon.

\subsection{Ansatz and equations} 
To investigate the inner structure of equal-spinning Myers-Perry BHs with scalar hair~\cite{Brihaye:2014nba}, we  consider a line element with a Schwarzschild-like radial variable:
 \begin{eqnarray}
\label{metric}
ds^2 & = & -b(r) dt^2 + \frac{dr^2}{f(r)}  + g(r) \left\{d\theta^2 + h(r)\left(\sin^2\theta \left[d\varphi_1 - 
W(r) dt\right]^2 + \cos^2\theta\left[d\varphi_2 -W(r)dt\right]^2\right)\right. \nonumber \\
&&\ \ \ \ \ \ \ \ \ \ \ \ \ \ \ \ \ \ \ \ \ \ \ \ \ \ \ \ \  \ \ \ + 
\left.\left[1-h(r)\right] \sin^2\theta \cos^2\theta (d\varphi_1 - d\varphi_2)^2\right\} \ .
\end{eqnarray}
Here the angular coordinates are a bipolar parameterization of $S^3$: $\theta$ runs from $0$ to $\pi/2$, while $\varphi_1,\varphi_2\in [0,2\pi[$. 
These space-times rotate in two orthogonal 2-planes ($\theta=0$ and $\theta=\pi/2$) and imposing equal angular momenta implies these two 2-planes can be interchanged; thus the spatial isometry
group is $U(2)$.

The metric above still leaves some gauge freedom: the diffeomorphism related to the definition of the radial variable $r$. For the numerical construction, we will fix this  freedom by choosing $g(r)=r^2$. We observe that this choice of coordinates is different from the one in~\cite{Brihaye:2014nba}.

The ansatz (\ref{metric}) for the metric is completed with an appropriate ansatz
for the scalar fields  originally proposed in  \cite{Hartmann:2010pm}
for the study of the boson star solutions of the same model:
\be
\label{scalar_ansatz}
           \Pi =  \e^{-i \omega t}  \phi(r)
     \left(      \begin{array}{c}
            \sin \theta \e^{i \varphi_1} \\ \cos \theta \e^{i \varphi_2}  
            \end{array}
            \right) \ ,
\ee
involving a harmonic time dependence with frequency $\omega$. 

The full ansatz (\ref{metric})-(\ref{scalar_ansatz})
leads to a system of five differential equations for the functions $b,f,h,W, \phi$. 
Choosing appropriate combinations of the Einstein equations, the
coupled system can be set in a form such that the 
equation for $f$ is a first order ordinary differential equation (ODE) 
while the other equations are second order ODEs.
This requires specifying nine conditions for a boundary value problem. The choice of these boundary conditions depends on the type of solution one wishes to discuss: a boson star, which is everywhere regular, including at the origin $r=0$, or a BH with an event horizon, say, at $r=r_H$.

\subsection{Boundary conditions}
For boson stars, regularity of the metric functions and of the scalar fields
 at the center of the soliton, $i.e.$ at $r=0$, requires the following conditions:
\be
   f(0) = 1 \ , \qquad \ b'(0) = 0 \ , \qquad h'(0) = 0 \ , \qquad   W'(0) = 0  \ , \qquad  \phi(0) = 0 \ .
\ee 
Henceforth, prime denotes the derivative with respect to $r$.
For BHs, on the other hand, for $r=r_H$ to be a regular horizon, 
the field equations  have to be solved with the following conditions:
\be
\label{regular_rh}
f(r_H) = 0 ,~ b(r_H) = 0,~W(r_H) = \omega,~
 G_1(b',h,h',W,W';\phi,\phi')|_{r=r_H} = 0 ,~G_2(b',h,h',W,W';\phi,\phi')|_{r=r_H} = 0.
\ee 
Observe, in particular, that  the angular velocity of the horizon $W(r_H)$ needs to equal the frequency $\omega$;
 this is the synchronization condition for this ansatz. 
In (\ref{regular_rh})
 $G_1, G_2$ represent two polynomials 
 which have to vanish in order to guarantee the regularity of the solutions at the horizon. 
 The explicit form of $G_1,G_2$ is involved and not illuminating,  so we do not write it explicitly.
   
 For the metric to be asymptotically flat,  the following conditions are imposed (for both boson stars and BHs):
 \be
       f(r \to \infty) = 1  \ ,\qquad   b(r \to \infty) = 1  \ , \qquad  h(r \to \infty) = 1 \ , \qquad 
       W(r \to \infty) = 0 \ , \qquad  \phi(r \to \infty) = 0 \ .
 \ee
 The nine boundary conditions are thus specified for both cases.

We remark that the scalar field's mass $\mu$ can be rescaled into the radial variable $r$ and
  Newton's constant can be rescaled into the scalar field. We use this freedom to set $\mu=1$ and $8 \pi G =1$,
 in the equations of motion. 
Accordingly, only the parameters $\omega$ and $r_H$ (the latter, only in the case of BHs)  have to be specified.

 \subsection{Physical quantities}
 The different solutions can be characterized by several physical quantities. The Hawking temperature $T_H$
 and area $A_H$ of the BHs can be estimated from the metric potentials at the horizon:
 \be
          T_H = \frac{1}{4 \pi} \sqrt{ f'(r_H) b'(r_H) } \ , \qquad  A_H = V_3 r_H^3 \sqrt{h(r_H)} \ , \qquad 
					{\rm where}~~~~V_3 \equiv 2 \pi^2 \ .
 \ee  
 The ADM mass and angular momentum can be extracted from the asymptotic decay of the fields $f(r)$ and $W(r)$,  respectively (see e.g. \cite{Brihaye:2010wx} for more details):
 \be
         M_{ADM} = - \frac{3 V_3}{16 \pi G} {\cal U} \  , \qquad J =  \frac{V_3}{8 \pi G} {\cal W}  \ ,
 \ee
 where $\cal U$ and $\cal W$ are read off from the metric functions:
 \be
       f(r) = 1 + \frac{\cal U}{r^2} + \mathcal{O}\left(\frac{1}{r^4}\right)  \ , \qquad W(r) = \frac{\cal W}{r^4} + \mathcal{O}\left(\frac{1}{r^6}\right) \ .
 \ee
 The conserved Noether charge $Q$ associated with the $U(1)$ symmetry of the Lagrangian can be computed as the integral
 \be
                     Q = - \int  \sqrt{- g} \ j^0 \ d^4 x \ 
                     = \ 4 \pi^2 \int \sqrt{\frac{bh}{f}} \frac{r^3}{b} [W(r)-\omega] \phi^2 dr  \ .
 \ee
 Finally, it will also be useful 
 to compute the Ricci scalar of the corresponding spacetime. This can be evaluated via the trace
 of the energy-momentum tensor. In the parameterisation (\ref{metric}), this trace takes the form
\be
         T_{\mu}^{\mu} = - 3 f \left(\frac{d \phi}{d r}\right)^2 - 5\mu^2\phi^2 + 
         3 \phi^2\left\{ \frac{1}{b}(W - \omega)^2 - \frac{1}{ r^2}(\frac{1}{h}+2)\right\} \ \ .
\ee

\section{Non-linear Analysis: results} 
\label{sec4}
 
 \subsection{Vacuum Myers-Perry solutions}
 The vacuum, $D=5$ equal spinning MP solution~\cite{Myers:1986un} with event horizon at $r=r_H$ and horizon angular velocity $\Omega_H$ can be written in the form~\eqref{metric} with:
 \bea
\label{MP}
 &&
f(r) = 1 - \frac{1}{1 - r_H^2 \Omega_H^2} \Bigl( \frac{r_H}{r} \Bigr)^2 + \frac{r_H^2 \Omega_H^2}{1 - r_H^2 \Omega_H^2} \Bigl( \frac{r_H}{r} \Bigr)^4 \ ,~~
 b(r) = 1 - \Bigl( \frac{r_H}{r} \Bigr)^2 \frac{1}{1 - \left[1 - (\frac{r_H}{r})^4\right]r_H^2 \Omega_H^2 } \ ,
\\
\nonumber 
&&
 h(r) =    1 +  \frac{r_H^2 \Omega_H^2}{1 - r_H^2 \Omega_H^2} \Bigl( \frac{r_H}{r} \Bigr)^4   \ ,~~
g(r)=r^2,
~~
 W(r) = \frac{\Omega_H}{1 - \left[1 - (\frac{r_H}{r})^4\right]r_H^2 \Omega_H^2   }  \Bigl( \frac{r_H}{r} \Bigr)^4 \ .
 \eea
 The generic MP solutions has an event horizon at $r=r_H$ and 
 presents an inner (Cauchy) horizon at $r=   r_-$ with $0 < r_- < r_H$, and a curvature singularity at $r=0$.
 The solution exists for $r \in ]0, \infty[$; the functions $b(r), W(r)$ 
 remain finite in the limit $r \to 0$ while $f(r), h(r)$ diverge in this limit.

 \subsection{Hairy Myers-Perry solutions and boson stars}
 When the scalar field is non-zero, the field equations do not admit, to the best of our knowledge, closed form soliton or BH solutions. Such solutions, can, however be constructed numerically. 
In particular, we have constructed BH solutions~\cite{Brihaye:2014nba} by using a numerical routine based on the collocation method of \cite{colsys}.  
 
 The domain of existence of the equal spinning MP hairy BHs is shown in Fig.~\ref{mass_omega_fig}, in an ADM mass $vs.$ frequency diagram. 
 \begin{figure}[h]
\begin{center}
\includegraphics[width=10cm]{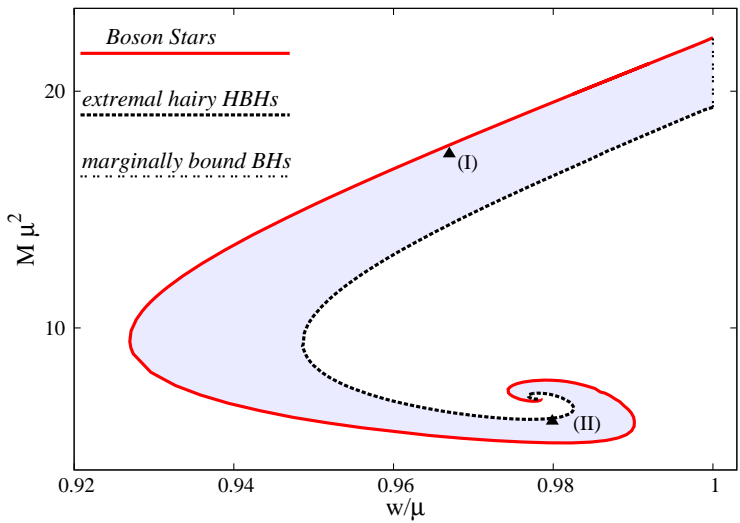}
\end{center}
\caption{ADM mass $vs.$ scalar field frequency for boson star solutions (red solid line) and for
 extremal hairy BHs  (black dotted line). Two specific solutions, both with $r_H=0.3$ and with $w_{\rm (I)}=0.967$ and $w_{\rm (II)}=0.98$, respectively are highlighted, that will be analysed in detail below.
\label{mass_omega_fig}
}
\end{figure}  
 As the figure shows, the hairy BHs exist for a range of frequencies, $0.927<w/\mu<1$ and it is bounded by the boson stars curve (red solid line) and the extremal BHs curve, for which $T_H=0$ (black dotted line). In the limit when $w=1$ the solutions approach locally a comparable vacuum MP solution, but not globally~\cite{Brihaye:2014nba}. 
In our numerical approach, the domain of existence of hairy MP solutions was spanned by lines of constant $r_H$, which, roughly, run parallel to the boson star line and end somewhere along the extremal BHs line. Thus, there can be two different solutions for the same values of  the input parameters
$w,r_H$. 
We have chosen, along the line with $r_H=0.3$, two illustrative solutions that will be examined in detail below, for their interior structure. These are marked as triangles in~Fig.~\ref{mass_omega_fig}, and have frequencies $w=0.967$ and $w=0.98$: configuration I along the first (top) branch and configuration II along the second (bottom) branch of an $r_H=0.3$ line. The latter is close to extremality.

We note, \textit{en passant}, that the value $b'(r_H)$ constitutes an appropriate parameter
to describe the solutions with a fixed $r_H$;
in particular, the temperature is a monotonic increasing function of $b'(r_H)$. 
Also, decreasing  the horizon radial coordinate $r_H$, the $\omega,M$ curve of the underlying hairy BHs 
progressively approaches the curve corresponding to the  boson stars. Finally, we remark that the BH sets  corresponding to a fixed $r_H$  bifurcate from the
family of extremal solutions.


\subsection{Moving inside the event horizon}
The extension of hairy BHs to the region inside the event horizon ($i.e.$ for $r \leq r_H$) is done in two steps. 
 Firstly, we obtain the exterior solution; thus, we integrate the system for $r \in [r_H, \infty]$,
 determining the values of the different functions on the horizon. Secondly, we use this set 
 as initial data to integrate inwards, on an interval  $[r_I,r_H]$ by decreasing progressively $r_I$. 
 
 As we shall see, the generic behaviour is that the curvature scalar becomes very large at non-vanishing values of the radial coordinate $r$, indicating an essential singularity forms. There is no strict significance of this radial coordinate we are using. From the metric~\eqref{metric}, however, it can be observed that the circumferencial radius of the orbits of the azimuthal Killing vector field ${\bf m}_1=\partial_{\varphi_1}$  (${\bf m}_2=\partial_{\varphi_2}$), along the plane on which the orbits of ${\bf m}_2=\partial_{\varphi_2}$ (${\bf m}_1=\partial_{\varphi_1}$) have vanishing size, is $r\sqrt{h(r)}$ (after gauge fixing). As we shall see, $\sqrt{h(r)}$ will be always of order unity. As such, our radial coordinate can still be roughly interpreted as a circumferential radius, and thus has a rough geometrical significance.
 
\subsubsection{Illustrative case (I)}

As a first illustrative example, we present in Fig.~\ref{functions_fig} (left panels) the results for a hairy BH relatively close to a boson star -- configuration (I) in  Fig. \ref{mass_omega_fig}, corresponding to a (first branch) solution with $w=\Omega_H = 0.967$, $r_H = 0.3$.
As exhibited in the top left panel, the numerical results strongly suggest  that the metric functions $f(r), b(r)$  vanish at a non zero value of $r$, denoted $r= r_S$; we were able to reach typically $f(r) \sim 10^{-6}$. The metric functions $h(r),W(r)$, on the other hand, remain finite and non-zero for $r \to r_S$ (left middle panel). The scalar field, its radial derivative and the Ricci scalar, however, diverge as $r_S$ is approached, thus indicating the formation of a curvature singularity (left bottom panel). This is precisely the behaviour we would have expected, from the considerations in the Introduction and Section 1, since the condition $W(r_S)= \omega $ cannot be met at a would-be inner horizon. What we observe is that the Ricci scalar $R$ diverges as $r\rightarrow r_S$ and the hairy BH  cannot be extended for $r < r_S$. The precise determination of $r_S$ is challenging but we find $r_S \approx 0.1047$.  As a comparison, even if comparing the values of radial coordinates in the two spacetimes is without strict significance (but observe the comments above about $r$ being approximately a circumferential radius), the vacuum MP solution with the same $r_H$, $\Omega_H$ presents an inner horizon for $r_C^{MP} \approx 0.091$, a smaller value than that of $r_S$, and, of course, can be continued for $r \to 0$.  

The description we have made for this example is generic for the hairy BHs close the the boson star limit. The corresponding spacetime geometries can be extended inside the event horizon up to some radial coordinate, $r=r_S$, before a Cauchy horizon is found, wherein an essential singularity is approached, rather than a smooth Cauchy horizon. Finally, we remark that, in this case, we did not observe oscillations in the scalar field. Since this is a far from extremal solution, such absence is in agreement with the left panel in Fig.~\ref{MPfig}. 
Then, the lesson from the test field analysis, is that we cannot exclude oscillations are present, further into the strong curvature region. Clarifying this issue (likely)  requires a different formulation of the numerical problem. In any case, the main trend -- absence of a smooth Cauchy horizon -- is clear, regardless of this detailed behaviour.

\begin{figure}[h!]
\begin{center}
{\includegraphics[width=8.4cm]{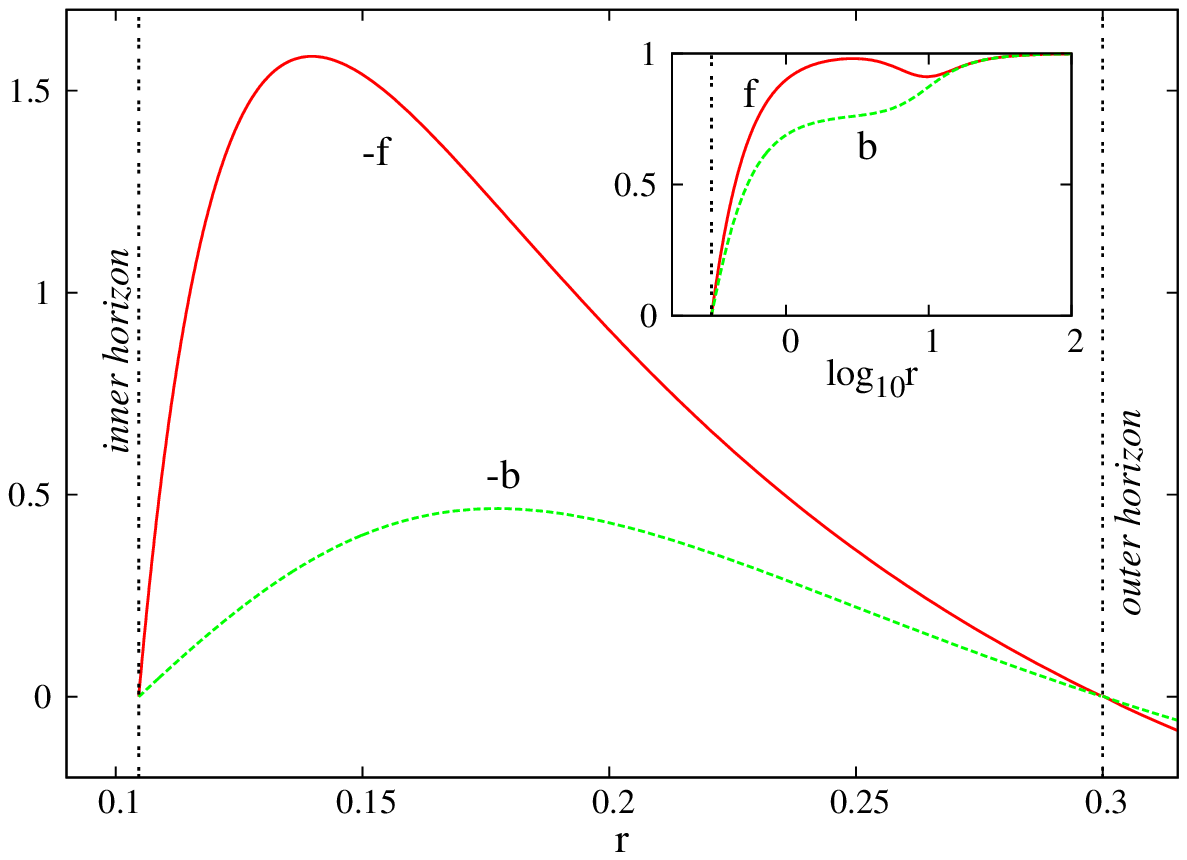}}
{\includegraphics[width=8.4cm]{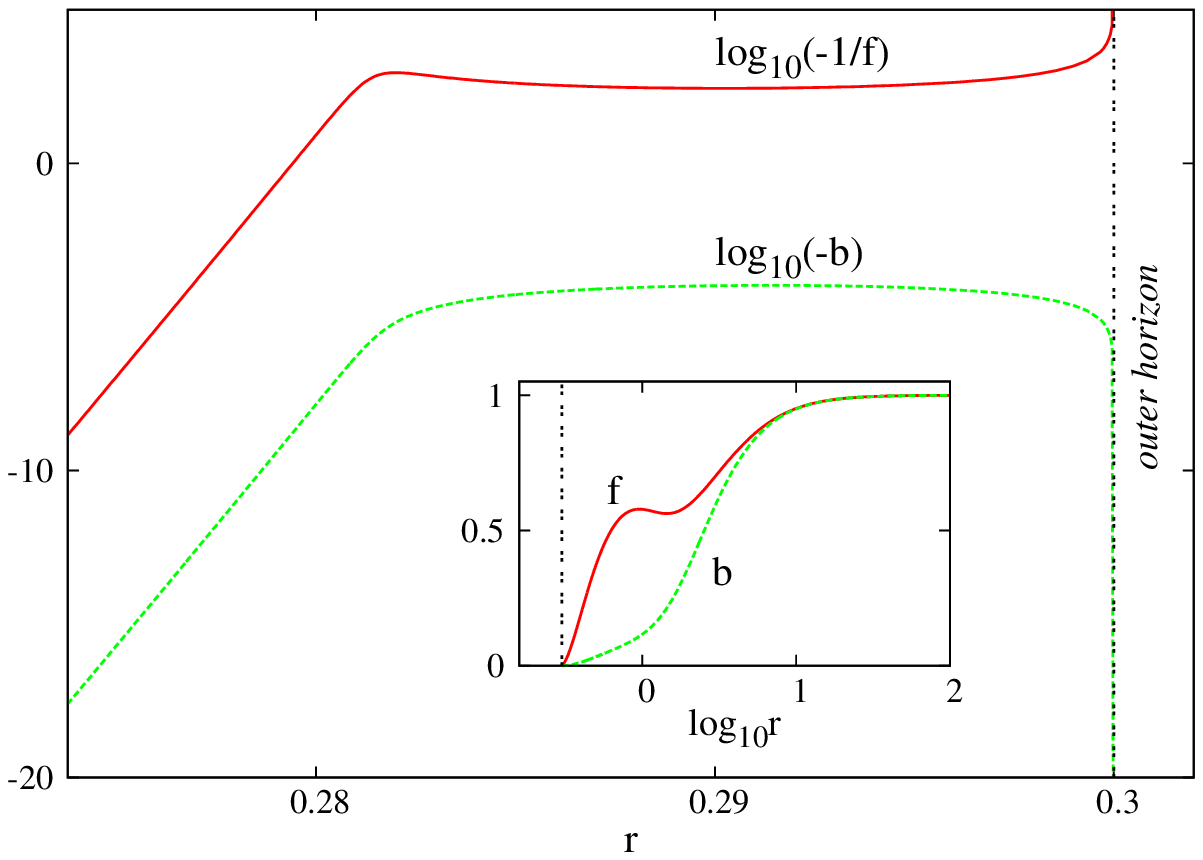}}
{\includegraphics[width=8.4cm]{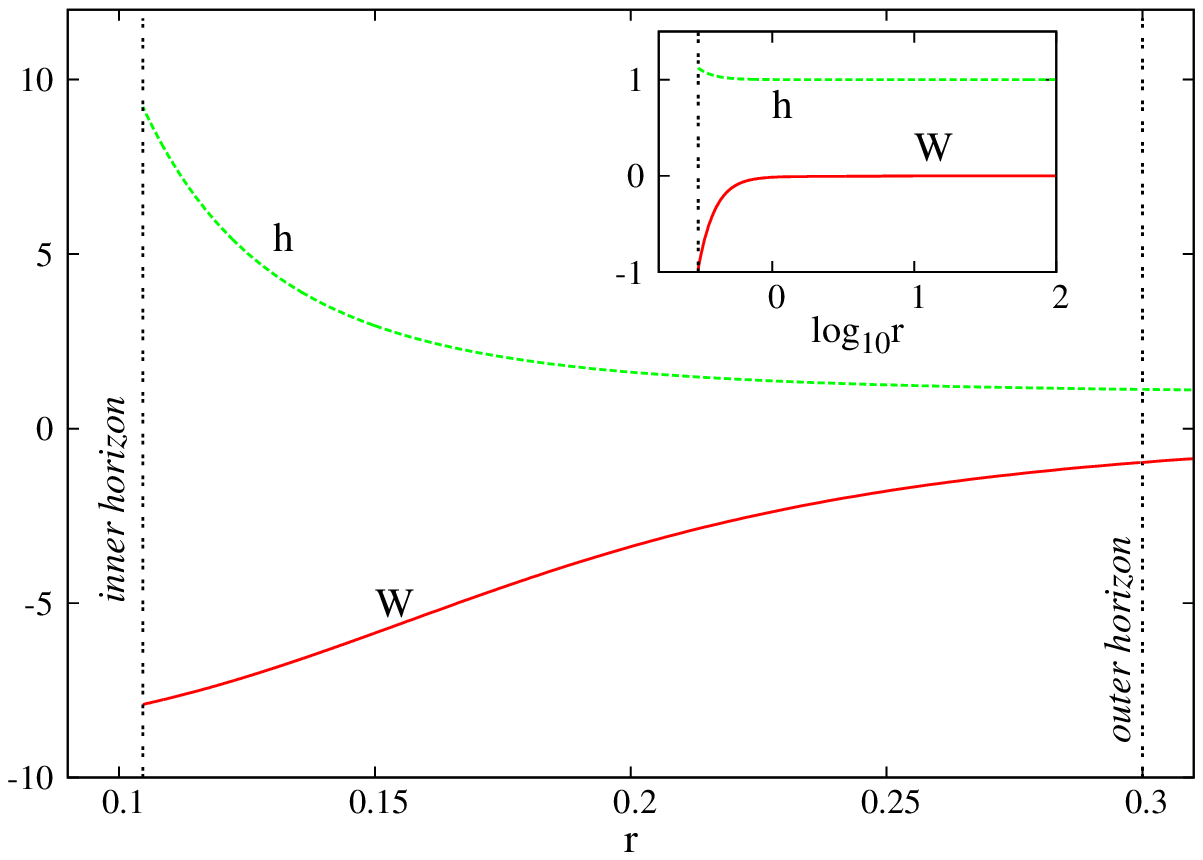}}
{\includegraphics[width=8.4cm]{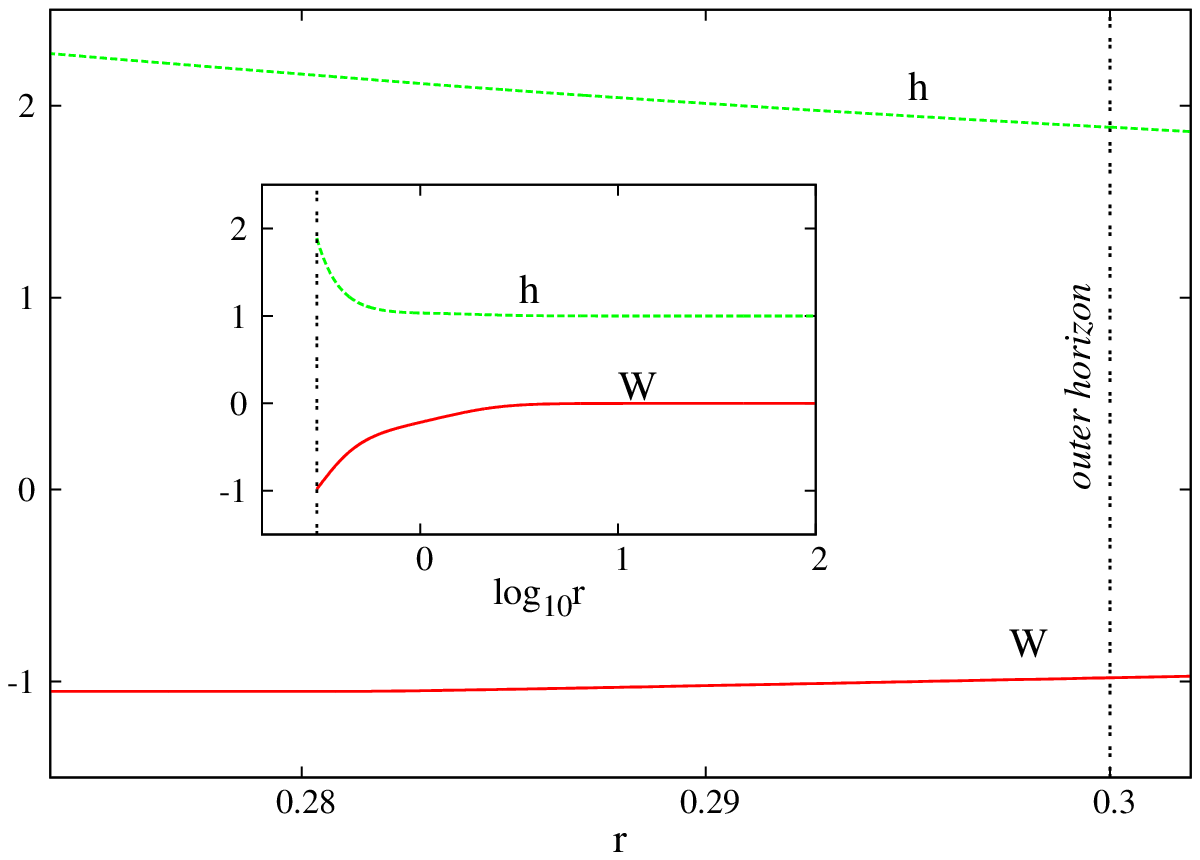}}
{\includegraphics[width=8.4cm]{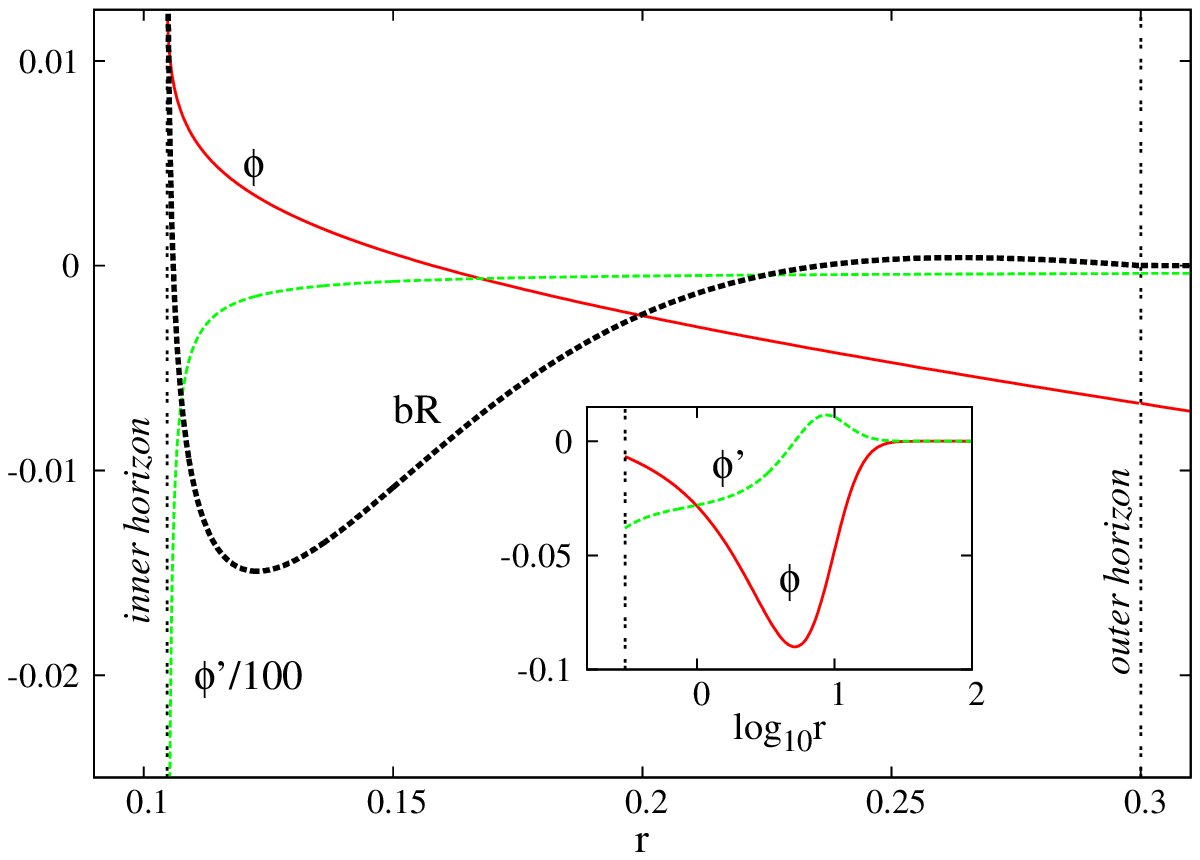}}
{\includegraphics[width=8.4cm]{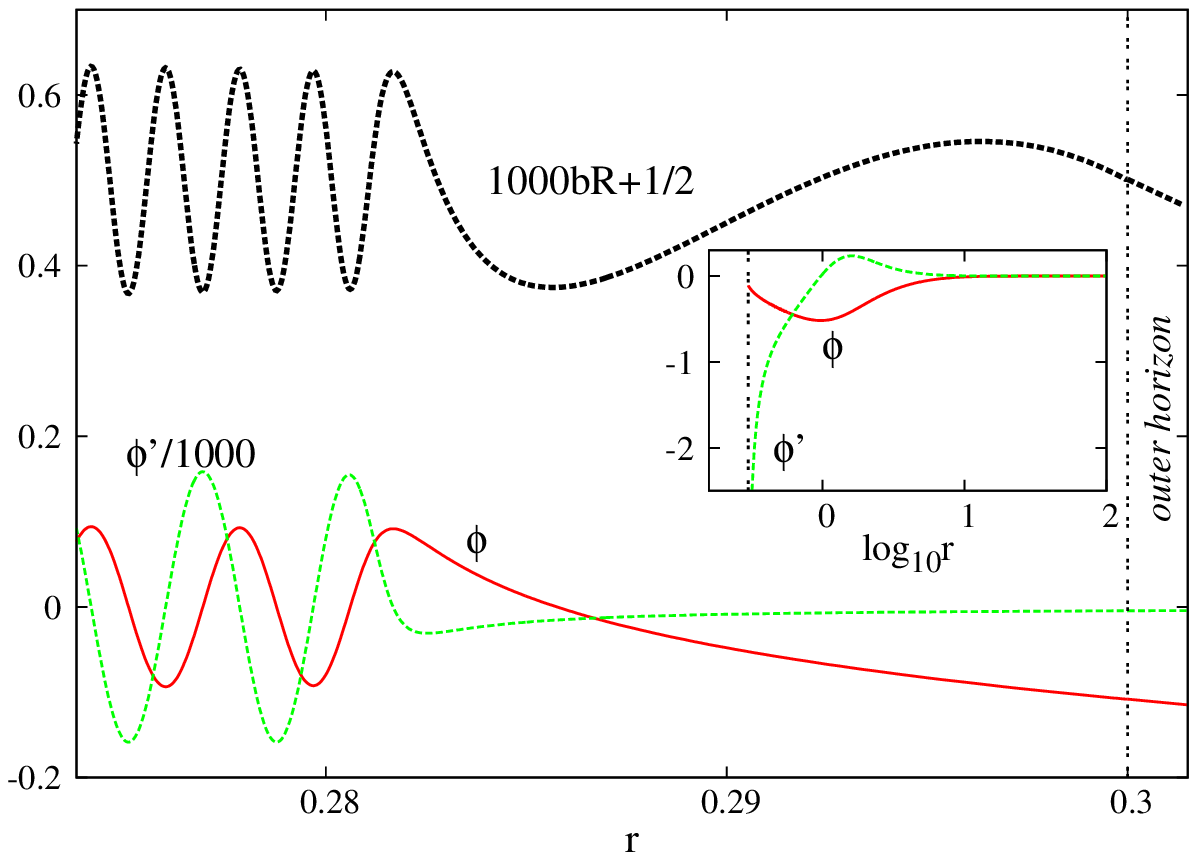}}
\end{center}
\caption{Behaviour of the various metric functions, the scalar field, its derivative and the curvature scalar for illustrative example I (left panels) and II (right panels).}
\label{functions_fig}
\end{figure}

\subsubsection{Illustrative case (II)}

As the second illustrative example, we present in Fig.~\ref{functions_fig} (right panels) 
the results for another hairy BH-- configuration (II) in  Fig. \ref{mass_omega_fig}, corresponding to a (second branch) solution with $w=\Omega_H = 0.98$, $r_H = 0.3$. 
For this case, the temperature of the hairy BH is very small and the solution is close to the extremal BHs line. 
We recall that, in the test field analysis, oscillations could be clearly seen in near-extremal solutions. This example will show they can also be seen in the non-linear analysis.

As can be observed from the top right panel, in this case, the metric function  $b(r)$ still become very small when integrating inwards;  $f(r)$, by contrast, starts to become exponentially large (in modulus) for $r<0.2815$. This exponential growth of $1/g_{rr}$ is precisely the trademark of mass inflation.\footnote{We want to emphasize that we are not claiming the existence of mass inflation, since we have not shown that there is a diverging quasi-local energy, such as the Misner-Sharp mass, which is only defined for spherically symmetric backgrounds. Our central claim is the absence of a smooth Cauchy horizon.} The behaviour of the metric functions $h(r),W(r)$, is not particularly distinct (middle right panel), and is even comparable to that observed for the previous case. The scalar field and its derivative on the other hand, start oscillating for $r<0.2815$, and the scalar curvature has exponentially growing oscillations. These features can be seen in the bottom right panel, where it should be noted that the Ricci scalar is being multiplied by the exponentially decreasing function $b(r)$. As a comparison, with the already mentioned limitations, we observe that the radial coordinate of the Cauchy horizon of the vacuum MP solution with the same $\Omega_H, r_H$ is $r_C^{MP}=0.092$

These features fit well with the expectations from the discussion in the Introduction and in Section I. Approaching a would-be Cauchy horizon, the scalar field, and its derivative, start oscillating. The corresponding (presumably) large gradients source large curvatures, with the behaviour of the Ricci scalar, nevertheless accompanying the pattern of the scalar field and developing itself oscillations with an exponential envelope. Simultaneously, the exponential growth of $1/f$ suggest a mass inflation phenomenon. The end-product seems to be, again that these solutions develop a curvature singularity and not a smooth Cauchy horizon.
 

 \section{Discussion}
 \label{sec5}
 
 In this paper we have initiated the
analysis of the internal structure of rotating BHs with synchronized hair~\cite{Herdeiro:2014goa,Herdeiro:2015gia,Kleihaus:2015iea,Herdeiro:2015tia,Herdeiro:2016tmi}. 
Firstly we have argued, using linear theory, that imposing the synchronization condition on the outer horizon prevents strong spatial oscillations; but then, these oscillations cannot be avoided at the inner (Cauchy) horizon. 
We have exemplified this feature 
both for stationary scalar clouds around Kerr BHs and for similar clouds for an equal spinning $D=5$ MP BH (in a cavity), establishing a parallelism between the two cases. Next we have addressed the fully non-linear problem, examining the internal structure of synchronized hairy BHs. Due to technical advantages we have used the MP case as our case study; but we anticipate the conclusions are more generic and, in particular, apply also the four dimensional Kerr case. 
We have provided evidence that there is no smooth Cauchy horizon; instead, a curvature singularity should form. 
Establishing the true nature of this singularity (say, if the shear or expansion of a congruence of geodesics diverges) is beyond the scope of this work and requires other types of techniques. 
But it seems clear that tidal forces diverge and hence there is a physical singularity beyond which the spacetime cannot be extended in a physical sense. 
 
 Our analysis considered only non-extremal solutions. The extremal case will have to be considered separately. We notice that, for the latter case, the argument that the synchronization condition cannot hold for both horizons, does not apply. Thus, the extremal case is, likely, qualitatively different.

 \vspace{0.5cm} 
\noindent
\section*{Acknowledgements}
C. H. and E. R. acknowledge funding from the FCT-IF programme. This project has received funding from the European UnionÕs Horizon  2020 research and innovation programme under the Marie Sklodowska-Curie grant agreement No 690904, and by the CIDMA project UID/MAT/04106/2013.


%
%
%
%
%
%

\end{document}